\documentclass[10pt,conference,letterpaper]{IEEEtran}
\IEEEoverridecommandlockouts

\usepackage{amsmath,amssymb}
\usepackage{booktabs}
\usepackage{cite}
\usepackage{graphicx}
\usepackage{multirow}
\usepackage{xcolor}
\usepackage{url}
\usepackage{algorithm}
\usepackage{algpseudocode}
\usepackage{enumitem}

\graphicspath{{paper_figures/}}

\newcommand{\clip}{\operatorname{clip}}
\newcommand{\CRVI}{\mathrm{CRVI}}
\newcommand{\Vref}{V_{\mathrm{ref}}}
\newcommand{\qmax}{\overline{q}}
\usepackage{graphicx}
\usepackage{amsmath,amssymb,amsfonts}
\usepackage[hidelinks]{hyperref}
\def\BibTeX{{\rm B\kern-.05em{\sc i\kern-.025em b}\kern-.08em
    T\kern-.1667em\lower.7ex\hbox{E}\kern-.125emX}}

\usepackage{svg}
\usepackage{scalerel}
\usepackage{tikz}
\usetikzlibrary{svg.path}
\definecolor{orcidlogocol}{HTML}{A6CE39}
\tikzset{
  orcidlogo/.pic={
    \fill[orcidlogocol] svg{M256,128c0,70.7-57.3,128-128,128C57.3,256,0,198.7,0,128C0,57.3,57.3,0,128,0C198.7,0,256,57.3,256,128z};
    \fill[white] svg{M86.3,186.2H70.9V79.1h15.4v48.4V186.2z}
                 svg{M108.9,79.1h41.6c39.6,0,57,28.3,57,53.6c0,27.5-21.5,53.6-56.8,53.6h-41.8V79.1z M124.3,172.4h24.5c34.9,0,42.9-26.5,42.9-39.7c0-21.5-13.7-39.7-43.7-39.7h-23.7V172.4z}
                 svg{M88.7,56.8c0,5.5-4.5,10.1-10.1,10.1c-5.6,0-10.1-4.6-10.1-10.1c0-5.6,4.5-10.1,10.1-10.1C84.2,46.7,88.7,51.3,88.7,56.8z};
  }
}
\newcommand\orcidicon[1]{\href{https://orcid.org/#1}{\mbox{\scalerel*{
\begin{tikzpicture}[yscale=-1,transform shape]
\pic{orcidlogo};
\end{tikzpicture}
}{|}}}}

\title{DynaTrust-VVC: Directional Physics-Informed Trust-Based Detection and Mitigation for Cyber-Resilient Multi-Agent Volt--VAR Control}

\author{Md~Fazley~Rafy$^{\textsuperscript{\orcidicon{0000-0003-3057-9546}}}$\,,~\IEEEmembership{Graduate Student Member, IEEE}, Kamrul~Hasan$^{\textsuperscript{\orcidicon{0009-0002-5834-8395}}}$\,,~\IEEEmembership{Graduate Student Member, IEEE,}
\\~Anurag~K.~Srivastava$^{\textsuperscript{\orcidicon{0000-0003-3518-8018}}}$\,,~\IEEEmembership{Fellow,~IEEE}
\thanks{Authors are with the Lane Department of Computer Science and Electrical Engineering, West Virginia University, Morgantown, WV 26505, USA. The authors would like to acknowledge partial support from the U.S. Department of Energy (DOE) and the Defense Advanced Research Projects Agency (DARPA). The authors also thank Dr. Niloy Patari for his technical support.}}

\begin{document}
\maketitle

\begin{abstract}
Distributed Volt--VAR control relies on voltage measurements and is therefore vulnerable to false-data injection. Neighbor corroboration can distinguish an isolated corrupted measurement from a physical disturbance, but coordinated agents can falsely corroborate one another. This paper proposes DynaTrust-VVC, a cyber-resilient multi-agent Volt--VAR framework that assigns each incoming neighbor message a directional trust value based on the consistency between its reported voltage increment and a sensitivity-based prediction from active- and reactive-power changes. Trust falls immediately after a physics mismatch and recovers gradually after sustained consistency. DynaTrust-VVC further uses a physics-adaptive evidence threshold and counterfactual safe-voltage recovery-based mitigation. The proposed method was validated using a nonlinear IEEE 123-bus feeder with minute-resolution residential profiles and 27 matched replicates. DynaTrust-VVC identifies a coordinated three-bus attack one minute after onset in all 27 matched replicates, whereas an otherwise identical configuration with fixed neighbor trust identifies none. The composite resilience index also increases from 0.227 to 0.315, with no statistically significant increase in the benign false-alarm rate.
\end{abstract}


\section{Introduction}

High rooftop-PV penetration introduces rapid voltage fluctuations and bidirectional power flows that conventional distribution feeders were not originally designed to accommodate~\cite{antoniadou2017distributed}. Inverter-based Volt--VAR control (VVC) addresses these challenges through standardized local reactive-power dispatch~\cite{photovoltaics2018ieee,patari2021distributed}, while multi-agent implementations extend this capability by coordinating voltage regulation across multiple buses~\cite{wang2021multi,wang2020data,ardakan2026multi}. Although this distributed architecture improves scalability,
it also increases the system’s exposure to cyberattacks. False data injection (FDI) attacks can manipulate local voltage
measurements through bias, ramp, replay, intermittent, or noisy signals, causing controllers to make incorrect state dependent decisions \cite{liang2016review,majumder2023cyber}. Since Volt–VAR control directly
relies on measured voltage, compromised measurements result in incorrect reactive-power dispatch that can propagate voltage
deviations throughout electrically coupled buses \cite{liu2024enhancing}. An effective edge defense for distributed VVC must distinguish malicious measurement manipulation from legitimate physical disturbances, such as cloud-induced PV fluctuations or sudden load changes, while remaining lightweight enough for execution within each inverter control loop. Existing centralized detection methods, including load-flow verification approaches, introduce communication latency and create single points of failure that become increasingly problematic as feeder size grows~\cite{isozaki2015detection,sarker2025enabling,liu2024enhancing}. More broadly, most FDI detection techniques formulate attack detection as monitoring or state-estimation problems that operate independently of the control loop~\cite{musleh2019survey}. Likewise, learning-based multi-agent VVC methods primarily optimize voltage regulation without explicitly addressing attack detection, localization, or mitigation~\cite{wang2021multi,wang2020data,ardakan2026multi}. Resilient multi-agent secondary control has also been investigated for energy storage systems under denial-of-service attacks, but these approaches target different devices and threat models~\cite{chen2022multi}. More recently, language-model-based detection has been explored, although its computational requirements remain unsuitable for real-time inverter-level deployment~\cite{selim2024large}. Sensitivity-weighted neighbor corroboration distinguishes cyber events from physical disturbances because genuine disturbances produce consistent responses across electrically coupled buses~\cite{majumder2023cyber}. However, it implicitly assumes that neighboring agents are trustworthy. When multiple electrically coupled neighbors are compromised simultaneously, they corroborate one another's falsified measurements, causing coordinated attacks to be misclassified as legitimate physical disturbances.
To address this limitation, this paper proposes DynaTrust-VVC, which replaces static corroboration weights with directional, per-link trust states derived from the physical consistency between a neighbor's reported voltage change and the voltage change predicted from observed active- and reactive-power injections using feeder sensitivity matrices. Trust collapses immediately when a physics inconsistency is detected and recovers gradually only after sustained consistency, preventing compromised neighbors from reinforcing falsified measurements while preserving corroboration during legitimate physical events. The framework further integrates a physics-adaptive detection threshold and counterfactual safe-voltage recovery directly into the distributed VVC loop, allowing an agent to estimate the voltage expected without the corrupted measurement and adapt its reactive-power command using trusted power changes.

The contributions of this paper are summarized as follows.
\begin{itemize}[leftmargin=*,nosep]
\item A directional, asymmetric physics-trust mechanism that prevents
coordinated compromised neighbors from retaining full corroboration weight.
\item Physics-adaptive evidence gating and counterfactual safe-voltage
recovery within the same multi-agent VVC loop.
\item A matched 27-replicate evaluation, internal module ablation, and
sensitivity-mismatch study that quantify the contribution and boundary of
each DynaTrust component.
\end{itemize}

\begin{figure*}[!t]
\centering
\includegraphics[width=0.82\textwidth]{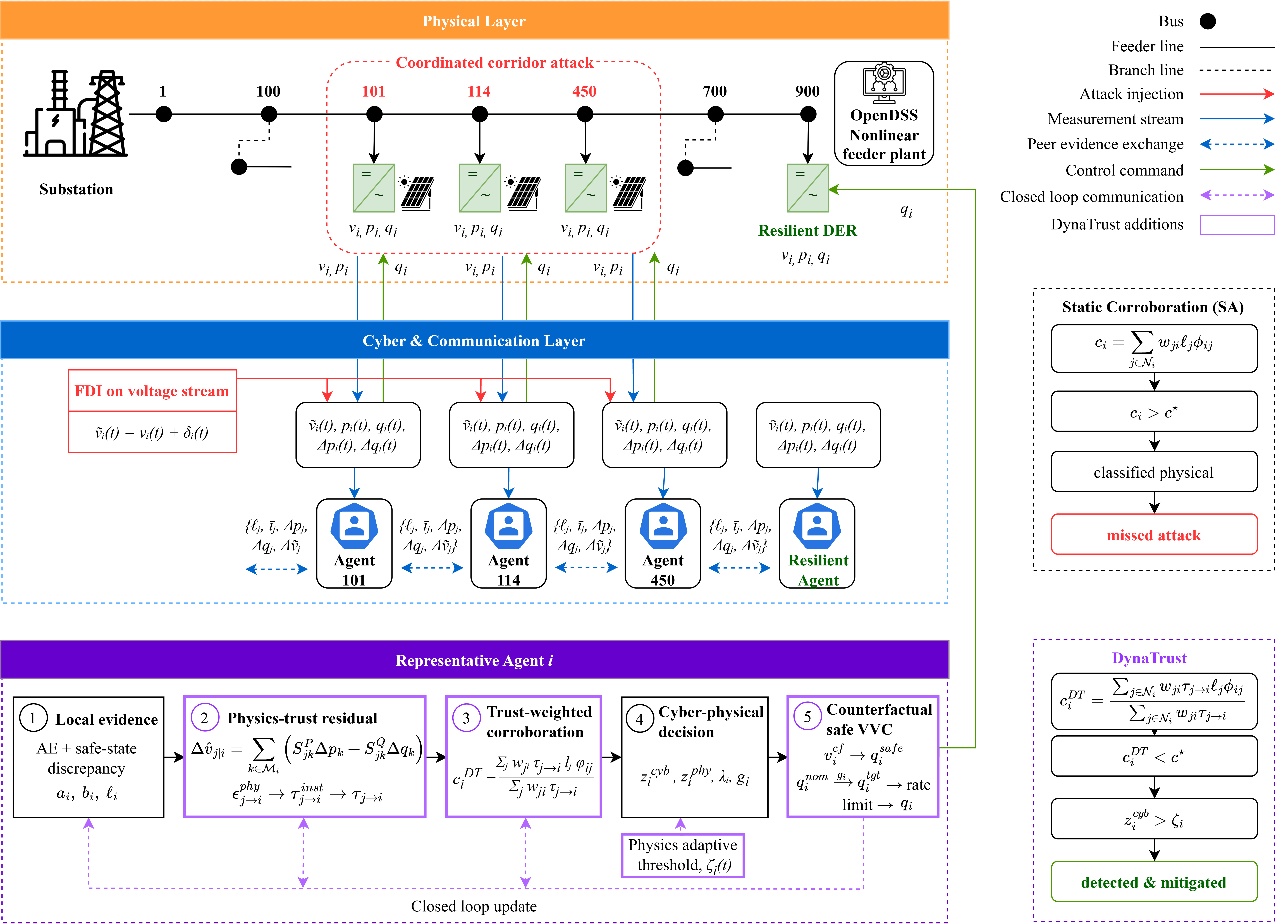}
\caption{Proposed DynaTrust-VVC framework for the multi-agent control pipeline, highlighting the dynamic physics-trust additions.}
\label{fig:arch}
\end{figure*}

\section{Proposed Methodology}\label{sec:method}

Consider a radial feeder with controllable inverter agents $\mathcal{A}=\{1,\ldots,N\}$. At minute $t$, agent $i$ receives voltage $\tilde v_i(t)$ and applies reactive command $q_i(t)$. The available reactive-power magnitude is denoted by $\overline{q}_i(t)$ and follows the inverter capability relation in Eq.~\eqref{eq:qcap}~\cite{photovoltaics2018ieee}.

\begin{equation}
\overline{q}_i(t)=
\sqrt{\max\!\left(s_i^2-p_i^2(t),\,0\right)}
\label{eq:qcap}
\end{equation}
Here, $s_i$ is the inverter apparent-power rating and $p_i(t)$ is the instantaneous active power. The nominal droop command is given in Eq.~\eqref{eq:droop}~\cite{photovoltaics2018ieee,patari2021distributed}.
\begin{equation}
q_i^0(t)=\clip\!\left(k_i\bigl(\Vref-\tilde v_i(t)\bigr),
-\qmax_i(t),\qmax_i(t)\right)
\label{eq:droop}
\end{equation}
The threat model is measurement-integrity corruption: $\tilde v_i(t)=v_i(t)+\delta_i(t)$, where $v_i(t)$ is the physical voltage and $\delta_i(t)$ can represent a bias, ramp, replay, intermittent, or noise attack~\cite{liang2016review}. Here, the feeder physics and actual power injections are not directly altered. For any time-varying signal $y_i(t)$, the one-minute increment is given in Eq.~\eqref{eq:increment}.
\begin{equation}
\Delta y_i(t)=y_i(t)-y_i(t-1)
\label{eq:increment}
\end{equation}
Each agent forms evidence from its controlled phases and transmits scalar
phase-aggregated quantities to its electrical neighbors. The local feature
vector is given in Eq.~\eqref{eq:feature}.
\begin{equation}
x_i(t)=\Bigl[\tilde v_i(t),\;
\tfrac{q_i(t)}{\qmax_i(t)},\;
\Delta\tilde v_i(t),\;
\tfrac{\Delta q_i(t)}{\qmax_i(t)}\Bigr]
\label{eq:feature}
\end{equation}
An autoencoder then reconstructs this $x_i(t)$ as $\hat{x}_i(t)$, and its normalized reconstruction score is given in Eq.~\eqref{eq:anom} \cite{sakurada2014anomaly}.

\begin{equation}
a_i(t)=\clip\!\left(
\frac{\lVert x_i(t)-\hat{x}_i(t)\rVert_2^2}{\theta_i},0,1\right)
\label{eq:anom}
\end{equation}

\begin{equation}
\iota_i(t)=\tilde v_i(t)-v_i^{\mathrm{safe}}(t)
\label{eq:discrepancy}
\end{equation}

\begin{equation}
b_i(t)=\clip\!\left(
\frac{\lVert\iota_i(t)\rVert_2}{\eta_i},0,1\right)
\label{eq:discrepancy_score}
\end{equation}
The validation-selected parameter $\theta_i>0$ is chosen as the reconstruction-error scale as given in Tab. \ref{tab:params}.
A second evidence channel compares the received voltage with the last trusted safe voltage $v_i^{\mathrm{safe}}(t)$, which remains frozen while cyber suspicion is latched. The safe-state voltage discrepancy and the corresponding bounded discrepancy score are given in Eq.~\eqref{eq:discrepancy} and Eq.~\eqref{eq:discrepancy_score}, respectively. Here, $\eta_i>0$ is a validation-selected discrepancy scale. The binary persistence variable $\gamma_i(t)$ is one only after the evidence candidate exceeds the trip level for two consecutive minutes with a consistent discrepancy sign. The transmitted local evidence is given in Eq.~\eqref{eq:local}.

\begin{equation}
\ell_i(t)=\gamma_i(t)\max\!\bigl(a_i(t),b_i(t)\bigr)
\label{eq:local}
\end{equation}

Agent $i$ exchanges $\ell_i(t)$, phase-averaged discrepancy $\bar{\iota}_i(t)$, normalized reactive-power increment $\bar d_i^q(t)$, and the increments $\Delta p_i(t)$, $\Delta q_i(t)$, and $\Delta\tilde v_i(t)$ only with the neighbor set $\mathcal{N}_i$. The compatibility factor between agents $i$ and $j$ is given in Eq.~\eqref{eq:agreement}.

\begin{equation}
\begin{aligned}
\phi_{ij}(t)
&=\exp\!\left(
-{|\bar{\iota}_i(t)-\bar{\iota}_j(t)|}/{\eta_i}\right)\\
&\quad{}\times
\exp\!\left(
-{|\bar d_i^q(t)-\bar d_j^q(t)|}/{\eta_i^q}\right)
\chi_{ij}(t)
\end{aligned}
\label{eq:agreement}
\end{equation}

Here, $\eta_i^q>0$ is the normalized reactive-increment scale. The sign-compatibility factor $\chi_{ij}(t)$ is one when nonzero discrepancies have the same sign, or when either discrepancy is zero; otherwise it is 0.35. Thus, $\phi_{ij}(t)\in[0,1]$ is large only when the two agents report compatible changes.
The fixed sensitivity weights satisfy $\sum_{j\in\mathcal{N}_i}w_{ji}=1$ and prioritize electrically influential neighbors~\cite{antoniadou2017distributed}. Then the common corroboration score is given in Eq.~\eqref{eq:corr}.
\begin{equation}
c_i(t)=\sum_{j\in\mathcal{N}_i}w_{ji}\ell_j(t)\phi_{ij}(t)
\label{eq:corr}
\end{equation}
A high corroboration score in Eq. \eqref{eq:corr} supports a common physical disturbance, whereas high local evidence with low corroboration supports localized corruption. Based on that, the cyber and physical scores are given in Eqs.~\eqref{eq:cyber} and \eqref{eq:physical}, respectively.
\begin{equation}
z_i^{\mathrm{cyb}}(t)=\ell_i(t)\bigl[1-c_i(t)\bigr]
\label{eq:cyber}
\end{equation}
\begin{equation}
z_i^{\mathrm{phy}}(t)=\ell_i(t)c_i(t)
\label{eq:physical}
\end{equation}
As given in Eq. \eqref{eq:gain}, the binary suspicion latch $\lambda_i(t)\in\{0,1\}$ activates after the persisted local evidence exceeds the trip level $\zeta_i(t)$ and releases when corroboration exceeds the physical-release level $c_\star$ or the latch horizon expires. Let $d_i(t)=\max\{z_i^{\mathrm{cyb}}(t),\ell_i(t)\}$, which is used to compute the mitigation gain in Eq.~\eqref{eq:gain}.
\begin{equation}
g_i(t)=\lambda_i(t)\mathbf{1}\!\left[c_i(t)<c_\star\right]
\clip\!\left(\frac{d_i(t)-\zeta_i(t)}{1-\zeta_i(t)},0,1\right)
\label{eq:gain}
\end{equation}
The diagnostic score $z_i^{\mathrm{cyb}}(t)$ determines whether a cyber event is supported by the evidence. The separate gain $g_i(t)$ determines how strongly the reactive-power command is moved toward its safe target. Then the nominal distributed target combines the droop command from Eq.~\eqref{eq:droop} with the bounded neighbor-consensus contribution $u_i^{\mathrm{cons}}(t)$, as given in Eq.~\eqref{eq:nominal_target}.
\begin{equation}
q_i^{\mathrm{nom}}(t)=
\clip\!\left(q_i^0(t)+u_i^{\mathrm{cons}}(t),
-\qmax_i(t),\qmax_i(t)\right)
\label{eq:nominal_target}
\end{equation}

The final target blends the nominal command and a safe command, as given in Eq.~\eqref{eq:trust}.
Here, $q_i^{\mathrm{safe}}(t)$ is the command calculated from the trusted safe-voltage estimate. When such an estimate is unavailable, the controller uses $q_i^{\mathrm{hold}}(t)$, the last reactive-power command stored before unconfirmed cyber suspicion. The common final command is rate-limited by the fraction $\rho_i$ of the available reactive capability and projected onto that capability interval in Eq.~\eqref{eq:final}.

\begin{equation}
q_i^{\mathrm{tgt}}(t)=
\bigl[1-g_i(t)\bigr]q_i^{\mathrm{nom}}(t)
+g_i(t)q_i^{\mathrm{safe}}(t)
\label{eq:trust}
\end{equation}


\begin{equation}
\begin{aligned}
q_i(t)
&=\clip\!\Bigg(q_i(t-1)
+\clip\!\Big(q_i^{\mathrm{tgt}}(t)-q_i(t-1),\\
&\qquad\quad-\rho_i\qmax_i(t),\rho_i\qmax_i(t)\Big),
-\qmax_i(t),\qmax_i(t)\Bigg)
\end{aligned}
\label{eq:final}
\end{equation}
Overall, Eqs. ~\eqref{eq:qcap}--\eqref{eq:final} define the common edge anomaly-detection, neighbor-evidence, and Volt--VAR control layer. The proposed DynaTrust-VVC framework extends this layer through directional physics-informed trust, physics-adaptive evidence gating, and a counterfactual safe-voltage recovery-based mitigation approach. The name DynaTrust refers to the time-varying trust assigned to each incoming neighbor message. 
\subsection{Proposed DynaTrust-VVC Extensions}\label{sec:dynatrust}
DynaTrust-VVC assigns a directional trust value $\tau_{j\to i}(t)\in[0,1]$ to each message received by agent $i$ from neighbor $j$. Let $\mathcal{M}_i=\{i\}\cup\mathcal{N}_i$. Agent $i$ predicts the voltage increment reported at neighbor $j$ using the first-order sensitivity model in Eq.~\eqref{eq:dv_hat}.
\begin{equation}
\Delta\hat{v}_{j|i}(t)=
\sum_{k\in\mathcal{M}_i}
\left[S^P_{jk}\Delta p_k(t)+S^Q_{jk}\Delta q_k(t)\right]
\label{eq:dv_hat}
\end{equation}
Here, $S^P_{jk}=\partial v_j/\partial p_k$ and $S^Q_{jk}=\partial v_j/\partial q_k$ are active- and reactive-power voltage sensitivities. They are obtained from the feeder model by centered finite differences, consistent with sensitivity-based distributed VVC \cite{antoniadou2017distributed}, \cite{patari2021distributed}. The normalized physics residual is given in Eq.~\eqref{eq:tau_resid}.
\begin{equation}
\varepsilon_{j\to i}^{\mathrm{phy}}(t)=
{\left|\Delta\tilde{v}_j(t)-\Delta\hat{v}_{j|i}(t)\right|}
/{\sigma_{\mathrm{phy}}}
\label{eq:tau_resid}
\end{equation}
The positive parameter $\sigma_{\mathrm{phy}}$ in Eq. \eqref{eq:tau_resid} is the allowable physics-residual scale. The instantaneous trust assigned to the message is then given in Eq.~\eqref{eq:tau_inst}.
\begin{equation}
\tau_{j\to i}^{\mathrm{inst}}(t)=
\exp\!\left(-{
\left(\varepsilon_{j\to i}^{\mathrm{phy}}(t)\right)^2}/{2}\right)
\label{eq:tau_inst}
\end{equation}
The stored trust state is updated asymmetrically. If the instantaneous trust is lower than the preceding stored trust, $\tau_{j\to i}^{\mathrm{inst}}(t)<\tau_{j\to i}(t-1)$, the message is discounted immediately according to Eq.~\eqref{eq:tau_drop}. 

\begin{subequations}
\label{eq:tau_asymm}

\begin{equation}
\tau_{j\to i}(t)=
\tau_{j\to i}^{\mathrm{inst}}(t)
\label{eq:tau_drop}
\end{equation}

Otherwise, when $\tau_{j\to i}^{\mathrm{inst}}(t)\geq\tau_{j\to i}(t-1)$, trust recovers gradually toward one according to Eq.~\eqref{eq:tau_recover}.

\begin{equation}
\begin{aligned}
\tau_{j\to i}(t)
&=\lambda_\tau\tau_{j\to i}(t-1)
 +(1-\lambda_\tau)
\end{aligned}
\label{eq:tau_recover}
\end{equation}

\end{subequations}

All trust states begin at one and $0<\lambda_\tau<1$ controls recovery. Thus, a newly inconsistent message loses influence immediately but regains influence only after sustained consistency. The trust-weighted corroboration used by DynaTrust-VVC is given in Eq.~\eqref{eq:corr_trust}.
\begin{equation}
c_i^{\mathrm{DT}}(t)=
\frac{\sum_{j\in\mathcal{N}_i}
w_{ji}\tau_{j\to i}(t)\ell_j(t)\phi_{ij}(t)}
{\sum_{j\in\mathcal{N}_i}w_{ji}\tau_{j\to i}(t)}
\label{eq:corr_trust}
\end{equation}
If the denominator in Eq.~\eqref{eq:corr_trust} is zero, the implementation sets $c_i^{\mathrm{DT}}(t)$ to zero. During DynaTrust-VVC operation, $c_i^{\mathrm{DT}}(t)$ replaces $c_i(t)$ in Eqs.~\eqref{eq:cyber}, \eqref{eq:physical}, and \eqref{eq:gain}. The expected voltage excursion used to adapt the evidence trip is given in  Eq.~\eqref{eq:omega}, and the adaptive evidence-trip level is given in Eq.~\eqref{eq:theta_adapt}.
\begin{small}
\begin{equation}
\begin{aligned}
\Omega_i(t)
&=\left|S^P_{ii}\Delta p_i(t)+S^Q_{ii}\Delta q_i(t)\right|\\
&\quad+\sum_{j\in\mathcal{N}_i}
\left[\left|S^P_{ij}\Delta p_j(t)\right|
+\left|S^Q_{ij}\Delta q_j(t)\right|\right]
\end{aligned}
\label{eq:omega}
\end{equation}
\end{small}
\begin{equation}
\zeta_i(t)=\clip\!\left(
\zeta\bigl[1+\kappa_\theta\Omega_i(t)\bigr],
\zeta,\zeta_{\max}\right)
\label{eq:theta_adapt}
\end{equation}
Here, $\zeta$ is the nominal trip level, $\kappa_\theta$ is the physics-adaptive gain, and $\zeta_{\max}<1$ bounds the trip level. However, this extension changes the evidence gate only; it does not alter the autoencoder threshold in Eq.~\eqref{eq:anom}.

When the cyber latch is active, DynaTrust-VVC propagates a counterfactual safe voltage, as estimated in Eq.~\eqref{eq:vcf}.
\begin{small}
\begin{equation}
\begin{aligned}
v_i^{\mathrm{cf}}(t)
&=\clip\!\Bigg(v_i^{\mathrm{cf}}(t-1)
+S^P_{ii}\Delta p_i(t)+S^Q_{ii}\Delta q_i(t)\\
&\quad+\sum_{j\in\mathcal{N}_i}\tau_{j\to i}(t)
\Bigl[S^P_{ij}\Delta p_j(t)+S^Q_{ij}\Delta q_j(t)\Bigr],
v^{\min},v^{\max}\Bigg)
\end{aligned}
\label{eq:vcf}
\end{equation}
\end{small}
The estimate is initialized from $v_i^{\mathrm{safe}}(t)$ at latch onset and reset when the latch is released. The DynaTrust-VVC safe command is the droop law evaluated at the counterfactual voltage, as given in Eq.~\eqref{eq:qsafe}.
\begin{equation}
q_i^{\mathrm{safe}}(t)=
q_i^0(t)\big|_{\tilde v_i(t)=v_i^{\mathrm{cf}}(t)}
\label{eq:qsafe}
\end{equation}
When trusted-peer support is insufficient, the implementation uses the last trusted command in place of $q_i^{\mathrm{safe}}(t)$ in Eq.~\eqref{eq:trust}. Algorithm~\ref{alg:agent} summarizes the resulting per-agent sequence from local evidence generation through trust update, gated recovery, and final reactive-power projection. Additionally, Table~\ref{tab:params} lists the parameters that affect the reported results. Validation days determine the detector and discrepancy scales, feeder physics determine the sensitivity matrices, and the remaining design constants are selected before test-set evaluation.

\begin{algorithm}[!b]
\caption{DynaTrust-VVC per-agent step at agent $i$}
\label{alg:agent}
\footnotesize
\begin{algorithmic}[1]
\State Receive $\tilde v_i(t)$; compute $\qmax_i(t)$ and cache $\Delta p_i,\Delta q_i$
\State Form $x_i(t)$ by \eqref{eq:feature}; score $a_i(t)$ by \eqref{eq:anom}; update $\ell_i(t)$ by \eqref{eq:local}
\State Broadcast $\{\ell_i,\bar{\iota}_i,\Delta p_i,\Delta q_i,\Delta\tilde v_i\}$ to $\mathcal{N}_i$
\State For each $j\in\mathcal{N}_i$, compute \eqref{eq:dv_hat} and update $\tau_{j\to i}$ by \eqref{eq:tau_asymm}
\State Compute $\zeta_i(t)$ by \eqref{eq:theta_adapt}, $c_i^{\mathrm{DT}}(t)$ by \eqref{eq:corr_trust}, and $g_i(t)$ by \eqref{eq:gain}
\If{$\lambda_i(t)=1$ and $g_i(t)>0$}
  \State Update $v_i^{\mathrm{cf}}(t)$ by \eqref{eq:vcf}; obtain $q_i^{\mathrm{safe}}(t)$ by \eqref{eq:qsafe}
\Else
  \State Set $q_i^{\mathrm{safe}}(t)\leftarrow q_i^{\mathrm{hold}}(t)$
\EndIf
\State Form $q_i^{\mathrm{tgt}}(t)$ by \eqref{eq:trust}; rate-limit and project by \eqref{eq:final}
\end{algorithmic}
\end{algorithm}

\begin{table}[t]
\centering
\caption{Parameters and provenance. V = selected on validation days,
P = physics-derived, D = fixed design constant.}
\label{tab:params}
\footnotesize
\begin{tabular}{llll}
\toprule
Symbol & Meaning & Value & Src.\\
\midrule
AE & autoencoder $4{-}8{-}3{-}8{-}4$, ReLU & 135 param. & D\\
$\theta_i^0$ & AE reconstruction threshold & 99th pct. & V\\
$\eta_i$ & safe-state discrepancy scale & 95th pct. & V\\
$|\mathcal{N}_i|$ & neighbors (strongest sensitivity) & 3 & D\\
$S^P,S^Q$ & voltage sensitivities & OpenDSS FD & P\\
$\sigma_{\mathrm{phy}}$ & physics-residual scale \eqref{eq:tau_resid} & 0.020 p.u. & D\\
$\lambda_\tau$ & trust recovery rate & 0.998 & D\\
$\kappa_\theta$ & threshold gain \eqref{eq:theta_adapt} & 5.0 & D\\
$\zeta$ & evidence trip / alarm level & 0.5 & D\\
$c_\star$ & physical-release level & 0.35 & D\\
--- & persistence & 2 min & D\\
--- & latch horizon & 45 min & D\\
$\rho_i$ & reactive ramp limit & 5\%/min & D\\
$\kappa$ & detection-delay decay constant & $15$ min & D \\
\bottomrule
\end{tabular}
\end{table}

\section{Experimental Design}
\label{sec:experiments}


The nonlinear IEEE 123-bus feeder is solved at one-minute resolution in
OpenDSS~\cite{kersting2001radial,opendss}. Ten controllable PV inverters use minute-resolution residential load and PV profiles from Pecan Street
Dataport~\cite{parson2015dataport}. Load is reconstructed as
\texttt{grid} $+$ \texttt{solar}; 281 complete common days are divided
chronologically into 168 training, 56 validation, and 57 test days. All defenses run as $N{=}10$ homogeneous agents on directed three-neighbor links, an arrangement previously demonstrated using distributed embedded controllers~\cite{rafy2026edge}.

Table~\ref{tab:scenarios} lists the primary scenarios used in the
quantitative comparison. All cyberattacks act from minute 360 through
minute 600.

\begin{table}[t]
\centering
\caption{Primary evaluation scenarios (minutes 360--600).}
\label{tab:scenarios}
\footnotesize
\begin{tabular}{@{}ll@{}}
\toprule
Class & Scenario \\
\midrule
Cyber & Bias, ramp, and replay at bus 114 \\
Cyber & Corridor bias: $-0.035$ p.u.\ at 101, 114, and 450 \\
Benign & Coupled PV drop \\
\bottomrule
\end{tabular}
\end{table}
\subsection{Configurations, Metrics, and Statistical Protocol}

The proposed configuration is DynaTrust-VVC (DT). The static-trust ablation (SA) uses the identical common layer but disables the three DynaTrust extensions by setting $\tau_{j\to i}(t)=1$, retaining $\zeta_i(t)=\zeta$, and setting $q_i^{\mathrm{safe}}(t)=q_i^{\mathrm{hold}}(t)$. The module chain further tests SA plus physics trust (PT), SA plus PT and adaptive thresholding (AT), and full DT, which adds counterfactual recovery (CR). A sensitivity-mismatch variant perturbs every agent-side $S^P_{jk}$ and $S^Q_{jk}$ entry by a fixed factor drawn from $U[0.8,1.2]$ while the feeder plant remains unchanged. Each run is compared with matched clean and unprotected trajectories having the same day and seed. Detection score $D$ uses the first-alarm latency $\ell_{\det}$ measured from attack onset. The detection-delay decay constant is fixed at $\kappa=15~\mathrm{min}$ before test-set evaluation.
\begin{equation}
D=\begin{cases}
\exp(-\ell_{\det}/\kappa), & \text{if an alarm is raised},\\
0, & \text{otherwise},
\end{cases}
\label{eq:D}
\end{equation}
Localization is the balanced average of recall and specificity, as given in Eq.~\eqref{eq:L}.
\begin{equation}
L=\tfrac{1}{2}\bigl(\mathrm{rec}+\mathrm{spec}\bigr);
\label{eq:L}
\end{equation}
Let $J$ be the weighted voltage--reactive-power tracking cost relative to
the clean run. The bounded mitigation score is given in Eq.~\eqref{eq:mit}.
\begin{equation}
M=\clip\!\left(1-\frac{J_{\mathrm{defended}}}
{J_{\mathrm{unprotected}}},0,1\right)
\label{eq:mit}
\end{equation}
The unbounded quantity $M_{\mathrm{raw}}=1-J_{\mathrm{defended}}/ J_{\mathrm{unprotected}}$ is additionally reported to expose degradation that clipping would otherwise hide. Availability $A$ is the mean fraction of feeder buses in $[0.95,1.05]$ p.u.\ during the attack window. Benign false-alarm rate (FAR) is the fraction of node-minutes satisfying $z^{\mathrm{cyb}}\ge\zeta$ during the benign-event window. The composite resilience index is the weighted geometric mean in Eq.~\eqref{eq:crvi}, motivated by the multi-dimensional nature of power system resilience~\cite{panteli2015grid}.
\begin{small}
\begin{equation}
\CRVI=\exp\!\left(\sum_{r}w_r\ln\!\left(\max(m_r,\epsilon)\right)\right)
\label{eq:crvi}
\end{equation}
\end{small}
Here, $m_r\in\{D,L,M,A\}$ and $w_r\in\{0.35,0.20,0.30,0.15\}$. All comparisons use 27 matched replicates, formed from nine random seeds across three held-out test days. Results are reported as mean $\pm$ standard deviation; principal comparisons use paired bootstrap 95\% confidence intervals with 20,000 resamples.

\section{Result Analysis}
\label{sec:results}

\begin{table}[t]
\centering
\caption{Comparison of the static-trust ablation (SA) and the proposed
DynaTrust-VVC framework (DT) over 27 matched replicates.}
\label{tab:head2head}
\footnotesize
\setlength{\tabcolsep}{3.1pt}
\begin{tabular}{@{}lcccccc@{}}
\toprule
& \multicolumn{2}{c}{$D$} & \multicolumn{2}{c}{$M_{\mathrm{raw}}$}
& \multicolumn{2}{c}{CRVI}\\
\cmidrule(lr){2-3}\cmidrule(lr){4-5}\cmidrule(lr){6-7}
Scenario & SA & DT & SA & DT & SA & DT\\
\midrule
Bias 114 & 0.936 & 0.936 & $+0.18$ & $+0.26$ & 0.566 & 0.621\\
Ramp 114 & 0.001 & 0.000 & $-0.05$ & $-0.03$ & 0.017 & 0.013\\
Replay 114 & 0.000 & 0.000 & $-0.10$ & $-0.24$ & 0.002 & 0.002\\
Coordinated corridor & 0.000 & \textbf{0.936} & $+0.02$ & $+0.03$ & 0.007 & \textbf{0.290}\\
\midrule
\multicolumn{7}{l}{Benign PV-drop FAR: SA $0.115\pm0.144$; DT $0.107\pm0.099$}\\
\bottomrule
\end{tabular}
\end{table}

\begin{table}[t]
\centering
\caption{Module ablation over the four benchmark attacks}
\label{tab:ablation}
\footnotesize
\setlength{\tabcolsep}{3.0pt}
\begin{tabular}{@{}lcccccc@{}}
\toprule
Method & $D$ & $L$ & $M$ & $A$ & CRVI & FAR\\
\midrule
No defense & 0.000 & 0.500 & 0.000 & 0.996 & 0.000 & 0.000\\
SA & 0.234 & 0.553 & 0.058 & 0.994 & 0.227 & 0.115\\
SA + PT & 0.468 & 0.557 & 0.077 & 0.994 & 0.315 & 0.111\\
SA + PT + AT & 0.468 & 0.558 & 0.076 & 0.994 & 0.315 & 0.120\\
\textbf{DT (full)} & \textbf{0.468} & 0.561 & 0.076 & 0.994 & \textbf{0.315} & \textbf{0.107}\\
\bottomrule
\end{tabular}
\end{table}

Table~\ref{tab:head2head} compares the static-trust ablation with the proposed framework. On the single-bus bias, both configurations detect the attack with a one-minute latency, while counterfactual recovery raises raw mitigation from $+0.18$ to $+0.26$. Under the coordinated-corridor attack, SA fails in all 27 replicates because mutually compromised neighbors retain full corroboration weight. DT discounts their messages through Eq.~\eqref{eq:corr_trust}, detects the attack one minute after onset in all cases, and raises the scenario CRVI from 0.007 to 0.290. The benign FAR change is $-0.008$ with paired 95\% CI $[-0.048,0.026]$. Across the four benchmark attacks, CRVI rises from 0.227 for SA to 0.315 for DT. The paired per-replicate change is $+0.101$ with bootstrap 95\% CI $[0.079,0.127]$ and is positive in all 27 replicates. The advantage remains positive under equal, detection-heavy, and mitigation-heavy CRVI weights.

\begin{figure}[!t]
\centering
\includegraphics[width=.9\columnwidth]{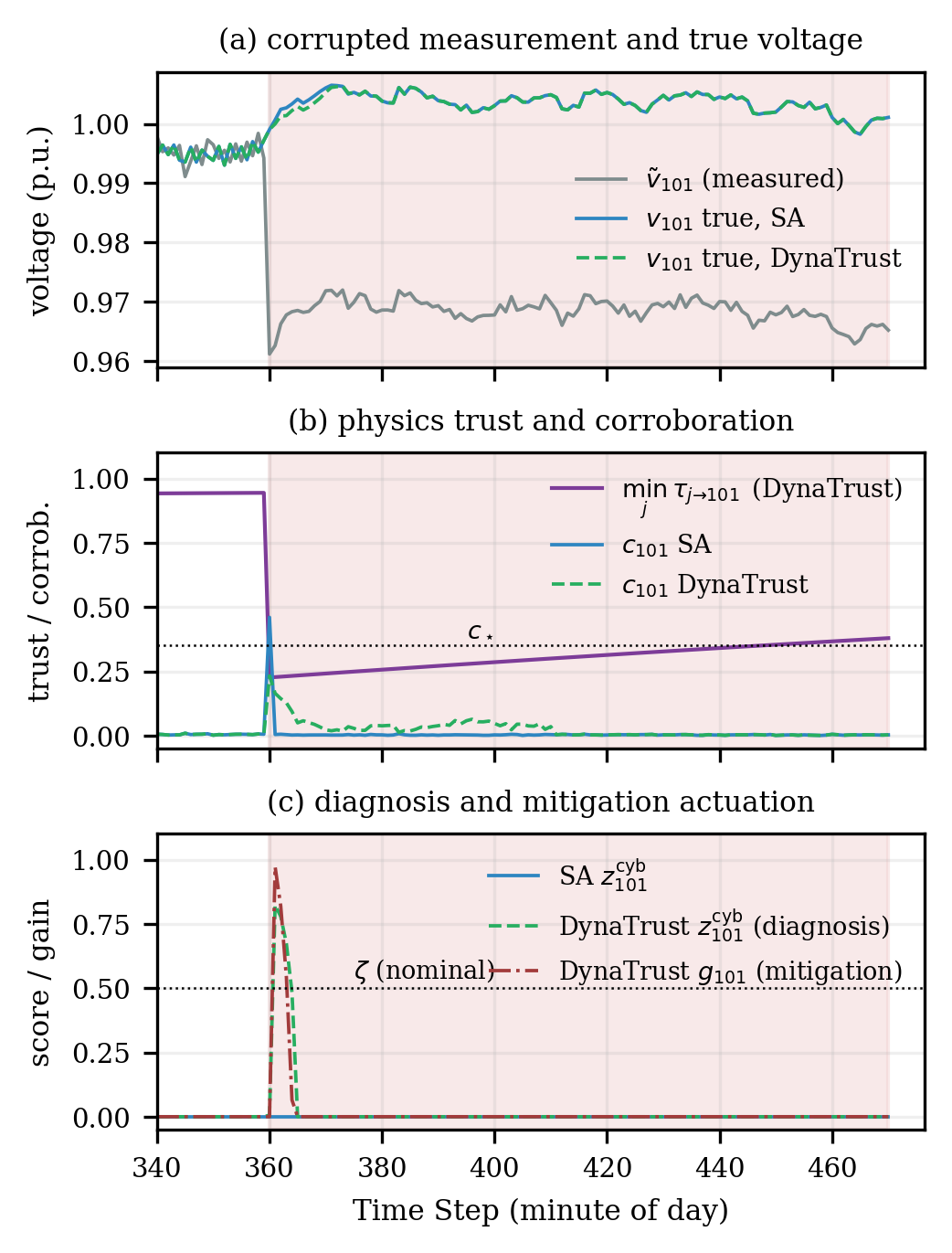}
\caption{Coordinated-corridor attack response at bus 101 detailing (a) voltage trajectories, (b) trust collapse preventing false physical corroboration, and (c) successful cyber score detection.}
\label{fig:case}
\end{figure}

Fig.~\ref{fig:case} shows a predefined seed-0, middle-test-day replicate. At minute 360, the three compromised buses report the same $-0.035$ p.u.\ step without a corresponding P--Q change. In SA, the three messages retain unit trust and corroborate one another; the cyber latch releases and no alarm is raised. In DT, the residual in Eq.~\eqref{eq:tau_resid} lowers directional trust, so the diagnostic score crosses the trip level one minute after onset. The separately gated mitigation gain then blends the command toward the counterfactual safe command. During the benign PV-drop event, the P--Q change explains the voltage variation, trust remains near one, and the event is classified as physical. Table~\ref{tab:ablation} identifies directional physics trust as the main source of coordinated-attack detection: pooled $D$ increases from 0.234 for SA to 0.468 after PT is enabled. AT preserves this gain during explainable electrical variation. Counterfactual recovery is most visible on the coordinated corridor, where raw mitigation increases from $+0.003$ for the PT+AT configuration to $+0.028$ for full DT. With a $\pm20\%$ perturbation of each agent-side sensitivity entry, coordinated detection remains 0.936 and raw mitigation remains $+0.028$; benign FAR increases from 0.107 to 0.125. However, ramp and replay remain difficult for SA and DT. Their incremental or replayed signatures are not sufficiently distinct from nominal load--PV variation over the 240-minute attack window. This operating boundary motivates sequence-aware detection and improved load-sensitivity estimation.

\section{Conclusions}
\label{sec:conclusion}
This paper proposed DynaTrust-VVC, a cyber-resilient multi-agent Volt--VAR control framework for inverter-rich distribution feeders. The framework combines local anomaly evidence and neighbor corroboration with directional physics-informed trust, adaptive evidence gating, and counterfactual safe-voltage recovery. These mechanisms enable each agent to assess whether neighboring messages are physically consistent before using them in its control decision. Results on the nonlinear IEEE 123-bus feeder with minute-resolution residential profiles show that DynaTrust-VVC detects the coordinated three-bus attack one minute after onset in all 27 matched replicates. The composite resilience index increases from 0.227 to 0.315 without a statistically significant increase in the benign false-alarm rate. The results further show that directional trust is particularly important when multiple electrically coupled agents are compromised simultaneously, while counterfactual recovery improves the resulting control response. The present evaluation is limited to the IEEE 123-bus feeder, and several controller design constants were fixed before test-set evaluation using the validation procedure summarized in Table~I. Although the $\pm20\%$ sensitivity-mismatch experiment preserves coordinated-attack detection, practical deployment will require sensitivity estimates that remain reliable as feeder topology and operating conditions change. Future work will examine parameter sensitivity, online sensitivity estimation, additional feeder models, and real-time control platforms.

\bibliographystyle{IEEEtran}
\bibliography{ref}

\end{document}